\documentclass[twocolumn,twocolappendix]{aastex631}
\usepackage{amsmath,amssymb,graphicx}
\usepackage{epsf,verbatim}
\usepackage{hyperref}
\usepackage{comment}
\usepackage{color}
\usepackage{slashed}
\usepackage{subfigure}
\usepackage{comment}
\usepackage{mdwlist, paralist}
\usepackage{rotating}
\usepackage{bm}
\usepackage{multirow}
\usepackage{Zallman,lettrine}

\newcommand{\beq}{\begin{equation}}
\newcommand{\eeq}{\end{equation}}
\newcommand{\bal}{\begin{align}}
\newcommand{\eal}{\end{align}}
\newcommand{\bit}{\begin{itemize}}
\newcommand{\eit}{\end{itemize}}
\newcommand{\ben}{\begin{enumerate}}
\newcommand{\een}{\end{enumerate}}

\renewcommand{\eqref}[1]{Eq.~(\ref{#1})}

\renewcommand{\t}{\tilde}

\newcommand{\f}{\frac}

\begin{document}

\title{Gravitational Waves from Primordial Black Hole Dark Matter Spikes}

\author[0000-0002-9048-2992]{Wei-Xiang Feng}
\email{wxfeng@mail.tsinghua.edu.cn; wfeng016@ucr.edu}
\affiliation{Department of Physics, Tsinghua University, Beijing 100084, China}
\affiliation{Department of Physics and Astronomy, University of California, Riverside, CA 92521, USA}

\author[0000-0001-5803-5490]{Simeon Bird}
\email{sbird@ucr.edu}
\affiliation{Department of Physics and Astronomy, University of California, Riverside, CA 92521, USA}

\author[0000-0002-8421-8597]{Hai-Bo Yu}
\email{haiboyu@ucr.edu}
\affiliation{Department of Physics and Astronomy, University of California, Riverside, CA 92521, USA}

\begin{abstract}
The origin of the binary black hole mergers observed by LIGO--Virgo--KAGRA remains an open question. 
We calculate the merger rate from primordial black holes (PBHs) within the density spike around supermassive black holes (SMBHs) at the centers of galaxies. We show that the merger rate within the spike is comparable to that within the wider dark matter halo. We also calculate the extreme mass ratio inspiral (EMRI) signal from PBHs hosted within the density spike spiralling into their host SMBHs due to gravitational wave emission. We predict that LISA  may detect $\sim10^4$ of these EMRIs with a signal-to-noise ratio threshold of 20 within a 4\,yr observation run, if all dark matter is made up of $\sim30{\rm\,M}_\odot$ PBHs. Uncertainties in our rates come from the uncertain mass fraction of PBHs within the dark matter spike, relative to the host central SMBHs, which defines the parameter space LISA can constrain.
\end{abstract}


\section{Introduction}
\lettrine[lines=5]{P}{rimordial black holes} (PBHs) have been proposed as a candidate for dark matter (DM), which makes up $85\%$ of the mass in the Universe\,\citep{Carr:1974nx,Meszaros:1974tb,Carr:1975qj}.
Observations from astrophysics and cosmology, notably gravitational microlensing of stars, supernovae, and
quasars, and accretion effects on the cosmic microwave background, have placed stringent constraints on the PBH DM abundance under the assumption of a monochromatic PBH mass spectrum 
\cite[see, e.g.,][]{Carr:2020gox,Bird:2022wvk}. 
The strongest constraints in the solar mass range are from microlensing surveys of stars in the Large Magellanic Cloud\,\citep{Macho:2000nvd,EROS-2:2006ryy,Wyrzykowski:2011aa,Blaineau:2022nhy,Mroz:2024mse}.

A new search channel for PBH DM has been opened by the LIGO--Virgo--KAGRA (LVK) detection of gravitational waves\,\cite[GWs;][]{KAGRA:2021duu}. Indeed, for reasonable PBH DM fractions the predicted merger rate may overlap the $17.9\textup{--}44\,{\rm Gpc^{-3}\,yr^{-1}}$ range estimated by LVK for binary black hole mergers of mass $\sim5\textup{--}100\,{\rm M}_\odot$\,\citep{Bird:2016dcv,Sasaki:2016jop,Raidal:2017mfl,Ali-Haimoud:2017rtz,Chen:2018czv,Liu:2018ess,Boehm:2020jwd,Hutsi:2020sol,Franciolini:2021tla,Jangra:2023mqp,Andres-Carcasona:2024wqk}.

The rate at which a population of PBHs produces observable gravitational-wave (GW) mergers is uncertain and widely discussed. \cite{Bird:2016dcv} suggested that PBH binaries forming via gravitational capture within low redshift halos could match the LVK merger rate if the PBH DM fraction were approximately unity. An alternative channel is the formation of binaries during the radiation-dominated era. Early estimates suggested a larger merger rate and thus implied that a smaller PBH DM fraction ($f_{\rm PBH}\sim10^{-3}$) was necessary to match the LVK results\,\citep{Sasaki:2016jop,Ali-Haimoud:2017rtz}. However, later analysis suggests that three-body interactions prevent the early binaries thus formed from surviving to the present day\,\citep{Raidal:2018bbj,Jedamzik:2020omx,Jedamzik:2020ypm}. Even with this consideration, the merger rate of the early binaries is still much greater than those through gravitational capture at later times\,\citep{Vaskonen:2019jpv,Raidal:2024bmm}, and the constraint is still as strong as $f_{\rm PBH}\sim10^{-3}$\,\citep{Andres-Carcasona:2024wqk}.

In this paper, we investigate a novel channel for GW signals: PBHs merging with supermassive black holes (SMBHs) and PBHs forming binaries in the density spike around SMBHs. DM around SMBHs at the centers of galaxies will develop a high-density region, known as a density spike, within the radius of influence of the SMBH. A DM density spike containing PBHs will have a high PBH density and thus an enhanced LIGO-band merger rate\,\citep{Nishikawa:2017chy,Fakhry:2023ggw}. This channel can impose even stronger constraints on PBHs than microlensing, as any PBHs in broader DM halos will eventually sink into the galactic center due to mass segregation. Assuming $f_{\rm PBH} = 1$ in the spike can still be consistent with microlensing constraints, which enforce $f_{\rm PBH} < 1$ in the wider halo.

Extreme mass ratio inspirals (EMRIs) are among the primary sources for the upcoming LISA mission\,\citep{Gair:2004iv,Babak:2017tow,Wang:2019kzb,Bonetti:2020jku,Li:2021pxf,LISA:2022yao,Colpi:2024xhw}. EMRIs could be produced between PBHs in the density spike and the central SMBHs. 
In this {work}, we estimate the rate of EMRIs from PBH density spikes and examine the correlation between the EMRI and PBH binary merger rates. The PBH merger rate within the spike is comparable to that within the wider DM halo. The dominant EMRI formation channel is not two-body relaxation, as assumed in the literature\,\citep{Amaro-Seoane:2007osp,Amaro-Seoane:2012lgq}, but GW emission. We show that LISA can place strong constraints on the PBH abundance in the spike.

\section{Merger rate estimation in DM spikes} 
Consider two unbound point masses $m_1$ and $m_2$ on an initial close encounter with initial relative velocity $v_{\rm rel}$. We assume the orbit is close to parabolic and consider the leading term in the limit eccentricity $e \to 1$. The distance of closest approach is then
\begin{equation}\label{eq:closest}
r_{\rm min}=\frac{\sigma v_{\rm rel}^2}{2\pi G(m_1+m_2)}
\end{equation}
with $G$ Newton's constant and $\sigma$ the scattering cross section.
The total energy of gravitational radiation released during the passage is
\begin{equation}\label{eq:GW_energy}
\delta E_{\rm GW}=\frac{8}{15}\frac{G^{7/2}}{c^5}\frac{(m_1+m_2)^{1/2}m_1^2m_2^2}{r_{\rm min}^{7/2}}g(e)
\end{equation}
where $c$ is the speed of light and $g(e)$ depends only on the eccentricity. The two masses are gravitationally bound and eventually merge if the energy radiated through GW is larger than their initial kinetic energy, i.e., $\delta E_{\rm GW}\gtrsim\frac{1}{2}\mu v_{\rm rel}^2$, where $\mu\equiv\frac{m_1m_2}{m_1+m_2}$ is the reduced mass. The merging cross section is then\,\citep{Quinlan:1989,Mouri:2002mc}
\begin{equation}\label{eq:merg_sigma}
\sigma_{\rm merg}
=2\pi\left(\frac{85\pi}{6\sqrt{2}}\right)^{2/7}\frac{G^2(m_1+m_2)^{10/7}m_1^{2/7}m_2^{2/7}}{c^{10/7}v_{\rm rel}^{18/7}},
\end{equation}
where we have used $g(e=1)=425\pi/32\sqrt{2}$\,\citep{Turner:1977}. In practice, $v_{\rm rel}$ can be approximated by the DM velocity dispersion in a given region.

Consider the following two cases:
\begin{itemize}
\item PBH--PBH binary:
taking the PBH mass $m_1=m_2=m$, the merging cross section for two PBHs of equal mass is then
\[
\sigma_{\rm merg}^{\rm PBH}=4\pi\left(\frac{85\pi}{3}\right)^{2/7}\frac{G^2m^2}{c^4}\left(\frac{v_{\rm rel}}{c}\right)^{-18/7}.
\]
\item PBH--SMBH inspiral:
taking the SMBH mass $m_1=M>10^4 m=10^4 m_2$ for EMRIs, the merging cross section is then
\[
\sigma_{\rm merg}^{\rm EMRI}
=2\pi\left(\frac{85\pi}{6\sqrt{2}}\right)^{2/7}\left(\frac{M}{m}\right)^{12/7}\frac{G^2m^2}{c^4}\left(\frac{v_{\rm rel}}{c}\right)^{-18/7}.
\]
\end{itemize}
Note that there is a $\left(M/m\right)^{12/7}$ enhancement factor compared to the equal mass merger.

Now we consider $30{\rm\,M}_\odot$ PBHs, which lie in the window $\sim20\textup{--}100{\rm\,M}_\odot$ favored by LVK observations\,\citep{Carr:2020gox}, and Sgr-A*-like SMBHs. A central SMBH of mass $M=4.3\times10^6{\rm\,M}_\odot$\,\citep{GRAVITY:2023avo} has a radius of influence $r_{\rm sp}\simeq 2GM/v^2\simeq2.2{\rm\,pc}$ assuming the stellar velocity dispersion $v\simeq v_{\rm rel}\simeq130{\rm\,km/s}$ (as 1D dispersion $\sigma_*=v/\sqrt{3}\simeq75{\rm\,km/s}$). As we will show later, the total mass of $30{\rm\,M}_\odot$ PBH DM residing in the spike $M_{\rm sp}$ is (at most) $10\%$ of the SMBH, i.e., $0.1M$. Then the averaged density is $\rho=0.1M/V$ within the spike volume $V=4\pi r_{\rm sp}^3/3$. The merger rate of PBH binaries is
\begin{align}\label{eq:pbh_rate_est}
\mathcal{R}&\simeq\left(1/2\right)V\left(\rho/m\right)^2\sigma_{\rm merg}^{\rm PBH}v \simeq1.27\times10^{-11}\nonumber\\
&\times\left(\frac{4.3\times10^6{\rm\,M}_\odot}{M}\right)\left(\frac{v}{130{\rm\,km/s}}\right)^{31/7}{\rm yr^{-1}}.
\end{align}
The merger rate of EMRIs is
\begin{align}\label{eq:emri_rate_est}
\Gamma\simeq&\left(\rho/m\right)\sigma_{\rm merg}^{\rm EMRI}v
\simeq4.55\times10^{-7}\left(\frac{4.3\times10^6{\rm\,M}_\odot}{M}\right)^{2/7}\nonumber\\
&\times\left(\frac{30{\rm\,M}_\odot}{m}\right)^{5/7}\left(\frac{v}{130{\rm\,km/s}}\right)^{31/7}{\rm yr^{-1}}.
\end{align}  
Given $M\propto v^4$ by the $M\textup{--}\sigma_*$ relation\,\citep{Ferrarese:2000se,Gebhardt:2000fk,Tremaine:2002js}, Eqs.\,\ref{eq:pbh_rate_est} and \ref{eq:emri_rate_est} imply that $\mathcal{R}\propto M^{3/28}$ is independent of the PBH mass and \emph{weakly} dependent on the SMBH mass compared to $\Gamma\propto M^{23/28}m^{-5/7}$. 
The EMRI rate is estimated to be 4 orders of magnitude larger than that of PBH binary mergers in Sgr-A*-like spikes, and the rate increases with the SMBH mass as $\Gamma/\mathcal{R}\propto M^{5/7}m^{-5/7}$. 
We will show below that the actual enhancement factor is about $\mathcal{O}(10^2)$, after taking into account the fact that the distribution of the PBHs is not uniform and their density is higher near the SMBH. 
In the next section, we provide a more concrete model for the profile of the density spike and  calculate the rates more accurately.

\section{Modeling the DM density spike} 
A DM density spike may develop within the radius of influence $r_{\rm sp}$ of an SMBH. Assuming the main halo follows a Navarro--Frenk--White (NFW) profile\,\citep{Navarro:1995iw}, we consider a density spike following a power-law density profile
\begin{equation}
\rho_{\rm spike}(r)=
\begin{cases}
\rho_{\rm sp}\left(r/r_{\rm sp}\right)^{-\alpha},\,2GM/c^2<r\leq r_{\rm sp}; \\
0,\quad r\leq2GM/c^2.
\end{cases}
\end{equation}
Here $\rho_{\rm sp}$ is the scale density of the spike, to be fixed, and $\alpha$ is the logarithmic slope with $1\leq\alpha<3$ depending on astrophysical effects\,\citep{Ullio:2001fb,Merritt:2002vj,Gnedin:2003rj}. Note that $\alpha=1$ corresponds to the NFW inner cusp in the absence of an SMBH and $\alpha\rightarrow3$ is the upper limit for a finite spike mass. A single-mass PBH population in the spike, which reaches a steady state via gravitational (Coulomb) scattering, will follow the Bahcall--Wolf power law $\alpha=7/4$\,\citep{Bahcall:1976aa,Alexander:1999yz,Shapiro:2014oha}. This is valid only when the spike is fully relaxed.
Here we set the lower radial boundary of the spike to be $2GM/c^2$. A more detailed calculation gives $8GM/c^2$\,\citep{Gondolo:1999ef} (or $4GM/c^2$ from a general relativistic treatment\,\citep{Sadeghian:2013laa,Ferrer:2017xwm}), but this makes a negligible difference to the total spike mass:
\begin{align}
M_{\rm sp}&=\frac{4\pi\rho_{\rm sp}r_{\rm sp}^3}{3-\alpha}\left[1-\left(2GM/c^2r_{\rm sp}\right)^{3-\alpha}\right]\nonumber\\
&\simeq4\pi\rho_{\rm sp}r_{\rm sp}^3/(3-\alpha)\quad ({\rm as}~r_{\rm sp}\gg 2GM/c^2).
\end{align}

In our numerical study, we will set the spike radius around the SMBH using the Bondi accretion radius, $r_{\rm sp}=2GM/v_{\rm dm}^2$, where $v_{\rm dm}$ is the velocity dispersion of DM at $r \gg r_{\rm sp}$. 
For an NFW profile, 
$\rho_{\rm NFW}(r)=\rho_s\left(r/r_s\right)^{-1}\left(1+r/r_s\right)^{-2}$, we set $v_{\rm dm}=\sqrt{3}\,\sigma_{\rm dm}=\sqrt{3GM_{\rm max}/R_{\rm max}}$, where $\sigma_{\rm dm}$ is the 1D velocity dispersion of DM and $M_{\rm max}=4\pi\rho_s r_s^3f(c_{\rm max})$, with $f(x)=\ln(1+x)-x/(1+x)$, is the mass within $R_{\rm max}=c_{\rm max}r_s=2.1626r_s$.
As DM halos are in dynamical equilibrium with their central galaxies\,\citep{Zahid:2018}, the DM velocity dispersion $v_{\rm dm}$ is approximately the stellar velocity dispersion $\sqrt{3}\,\sigma_*$.
The scale density $\rho_s$ and radius $r_s$ can be translated into the halo mass $M_{200}$ and concentration $c_{200}$, then we use the $M\textup{--}\sigma_*$ relation in~\cite{Gultekin:2009aa} and \cite{Kormendy:2013dxa} and concentration--mass ($c_{200}\textup{--}M_{200}$) relation\,\citep{Prada:2011aa,Loudas:2022ipe} to associate the central SMBHs to their host halos.

We require that $M_{\rm sp}$ from DM is smaller than the uncertainty in determining the dynamical mass $M$ of the SMBHs\,\citep{Gorchtein:2010xa,Lacroix:2016qpq} and set $M_{\rm sp}=\Delta M$. For Centaurus A, $\Delta\lesssim0.5$\,\citep{Cappellari:2008db}; for Sgr\,A*, $\Delta\lesssim0.08$\,\citep{Gillessen:2008qv}; and for M87, $\Delta\lesssim0.1$\,\citep{EventHorizonTelescope:2019ggy,EventHorizonTelescope:2022wkp}.  
A more systematic treatment matches the Bahcall--Wolf spike to the inner cusp of the DM halo profile to fix $\rho_{\rm sp}$ with given $\Delta$. We choose $\Delta\simeq0.1$ to fix $\rho_{\rm sp}$ for an SMBH of $10^9{\rm\,M}_\odot$, which in turn fixes the inner cusp to $\propto r^{-\gamma}$ with $\gamma=1.7$  (see Appendix~B).

We assume the velocity of PBHs in the spike follows a Maxwell--Boltzmann distribution with a cutoff velocity $v_{\rm cut}$:
\begin{equation}\label{eq:MB}
\mathcal{P}(v,\sigma_{\rm sp})=F_0\left[\exp\left(-\frac{v^2}{\sigma_{\rm sp}^2}\right)-\exp\left(-\frac{v_{\rm cut}^2}{\sigma_{\rm sp}^2}\right)\right],
\end{equation}
where $F_0=F_0(\sigma_{\rm sp})$ is the normalization factor and the 1D velocity dispersion of DM in the spike is assumed to be
\begin{equation}
\sigma_{\rm sp}=\sqrt{\frac{G\left(M_{\rm sp}+M\right)}{r_{\rm sp}}}=\frac{v_{\rm cut}}{\sqrt{2}}
\end{equation}
with the cutoff velocity $v_{\rm cut}$ determined by the escape velocity in the density spike. Thus, PBH DM is bound into the spike but might be unbound or loosely bound to the central SMBH, and the velocity-weighted cross section is then $\langle\sigma_{\rm merg} v\rangle=\int_0^{v_{\rm cut}}\mathcal{P}(v, \sigma_{\rm sp})\sigma_{\rm merg}v{\rm\,d}^3v$.

\section{EMRIs versus PBH binaries}
We calculate the rate of PBH mergers within the spike as 
\begin{equation}\label{eq:pbh_rate}
\mathcal{R}=4\pi\int_{2GM/c^2}^{r_{\rm sp}}\frac{1}{2}\left(\frac{\rho_{\rm spike}(r)}{m}\right)^2\langle\sigma_{\rm merg}^{\rm PBH} v\rangle r^2{\rm\,d}r,
\end{equation}
while the EMRI merger rate is
\begin{equation}\label{eq:emri_rate}
\Gamma=\left(\frac{M_{\rm sp}/m}{4\pi r_{\rm sp}^3/3}\right)\langle\sigma_{\rm merg}^{\rm EMRI} v\rangle.
\end{equation}
Both EMRI and PBH mergers depend on central SMBH and halo properties in the same way,
\begin{equation}
\left\{\begin{array}{lr}
\mathcal{R}\\
\Gamma
\end{array}\right\}
\propto\frac{G^2\rho_s}{c^3}
\frac{M_{200}^2}{M}\frac{D(\sigma_{\rm sp})}{f^2(c_{200})}\frac{f^3(c_{\rm max})}{c_{\rm max}^3},
\end{equation}
up to different enhancement coefficients, $\mathcal{C}_{\rm PBH}$ and $\mathcal{C}_{\rm EMRI}$, respectively (see Appendix C and \cite{Feng:2023jvn} for derivation). 
Here $D(\sigma_{\rm sp})$ is given by
\begin{equation}
D(\sigma_{\rm sp})=\int_0^{v_{\rm cut}}c^3\mathcal{P}(v,\sigma_{\rm sp})\left(\frac{v}{c}\right)^{3/7}{\rm d}\left(\frac{v}{c}\right)
\end{equation}
using Eq.\,\ref{eq:MB}.
As expected, $\Gamma\propto\mathcal{C}_{\rm EMRI}/M\sim M^{-2/7}m^{-5/7}\Delta$ agrees with Eq.\,\ref{eq:emri_rate_est}. However, $\mathcal{R}\propto\mathcal{C}_{\rm PBH}/M\sim[(3-\alpha)^2\Delta^2/(2\alpha-3)](c^2/v_{\rm dm}^2)^{2\alpha-3}M^{-1}$ differs from Eq.\,\ref{eq:pbh_rate_est} by an extra enhancement $(c^2/v_{\rm dm}^2)^{2\alpha-3}$.  This comes from the integration lower bound in Eq.\,\ref{eq:pbh_rate} and is due to much higher density near the SMBH. Therefore, $\Gamma/\mathcal{R}\sim\mathcal{C}_{\rm EMRI}/\mathcal{C}_{\rm PBH}\sim[M^{5/7}m^{-5/7}(2\alpha-3)/\Delta(3-\alpha)^2](v_{\rm dm}^2/c^2)^{2\alpha-3}\sim\mathcal{O}(10^2)$, only 2 orders of magnitude larger, in contrast to our initial estimate. We also note that the merger rates are \emph{sensitive} to the spike fraction $\Delta$, where $\Gamma\propto\Delta$; whereas $\mathcal{R}\propto\Delta^2$. 

\begin{figure}[t]
\centering
   \includegraphics[width=0.47\textwidth]{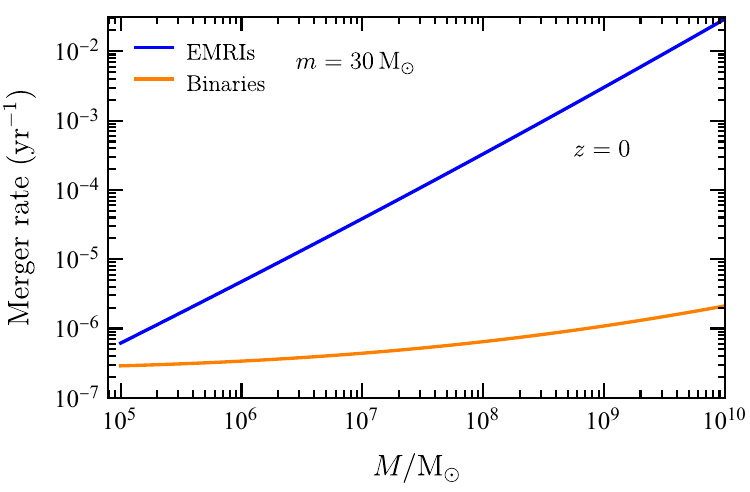}
   \caption{Merger rates of EMRI ({blue}) and PBH binary ({orange}) per halo as a function of the SMBH mass at redshift $z=0$, assuming the density spike is fully made of PBHs.
   }
   \label{fig:rate}
\end{figure}

Fig.\,\ref{fig:rate} shows the EMRI (blue) and binary (orange) merger rates driven by GW emission. The expected merger rates are comparable $\mathcal{R}\sim\Gamma\sim10^{-7}\textup{--}10^{-6}{\rm\,yr^{-1}}$ for SMBHs of $\sim10^5{\rm\,M}_\odot$; while the EMRI merger is more pronounced in higher-mass SMBHs, reaching $\sim10^{-2}{\rm\,yr^{-1}}$ at $\sim10^{10}{\rm\,M}_\odot$, which is 4 orders of magnitude larger compared to the PBH binary merger. This enhancement is mainly due to the factor $(M/m)^{5/7}$ with $M/m>10^4$, while the spike fraction $\Delta$ increases by only an order of magnitude.
For Sgr-A*-like SMBHs, particularly, we find $\mathcal{R}\simeq3.9\times10^{-7}{\rm\,yr^{-1}}$ is comparable to late-formed binaries in the wider DM halo with $f_{\rm PBH}\sim1$\,\citep{Bird:2016dcv,Raidal:2017mfl,Ali-Haimoud:2017rtz}; and $\Gamma\simeq1.8\times10^{-5}{\rm\,yr^{-1}}$ is predominant over the estimate of EMRI rates assuming two-body relaxation\,\citep{Hopman:2005vr,Hopman:2006xn,Rom:2024nso}.

%%%
%%%

As discussed previously, the calculation of merger rates shown in Fig.\,\ref{fig:rate} assumes the spike density follows the Bahcall--Wolf profile ($\alpha=7/4$)\,\citep{Bahcall:1976aa}. This is based on the assumption that the PBHs in the spike are relaxed, and the (two-body) relaxation time should be less than a Hubble time. The relaxation time is $\sim v^3/G^2\rho\,m\ln\Lambda$\,\citep{Merritt:2013de}
with the spike averaged density $\rho=3M_{\rm sp}/4\pi r_{\rm sp}^3\gtrsim3\Delta v^6/32\pi G^3M^2$ and $\ln\Lambda\sim\mathcal{O}(10)$ the usual Coulomb logarithm. 
Using Eq.\,\ref{eq:MB} for root-mean-square velocity $v_{\rm rms}\equiv\sqrt{\langle v^2\rangle}$, we estimate the relaxation time for the spike as
\begin{align}
t_{\rm relx}&\simeq\frac{GM^2}{mv_{\rm rms}^3\Delta\ln\Lambda}\simeq9.34{\rm\,Gyr}\left(\frac{0.039}{\Delta}\right)\left(\frac{8.3}{\ln\Lambda}\right)\nonumber\\
\times&\left(\frac{M}{4.3\times10^6{\rm\,M}_\odot}\right)^2\left(\frac{30{\rm\,M}_\odot}{m}\right)\left(\frac{95{\rm\,km/s}}{v_{\rm rms}}\right)^3,
\end{align}
which is less than a Hubble time ($t_H\simeq14{\rm\,Gyr}$) only for SMBHs $\lesssim10^6{\rm\,M}_\odot$. The relaxation time may be lowered by the presence of stellar objects (other than PBHs) in the spike, with $\Delta=M_{\rm sp}/M\sim\mathcal{O}(1)$. This suggests that spikes could be relaxed in a Hubble time for SMBHs of $M \lesssim 10^7{\rm\,M}_\odot$\,\citep{OLeary:2008myb,LISA:2022yao}. Larger SMBHs may not have developed the Bahcall--Wolf profile in the spike and may not have a Maxwell--Boltzmann velocity distribution. Although this might alter our predicted merger rates, LISA is likely not sensitive to EMRIs onto SMBHs more massive than $\sim10^7{\rm\,M}_\odot$, and so our estimates for SMBHs $\lesssim10^7{\rm\,M}_\odot$ can still constrain PBHs in the spike.

Note the rates shown in Fig.\,\ref{fig:rate} are conservative as they consider only mergers formed by GW emission. In reality, the SMBHs will be surrounded by nuclear star clusters\,\citep{Zhou:2024duc,Mukherjee:2024krx}. PBHs may experience dynamical friction when passing through the stars colocated around SMBHs with the spike. In particular, for relaxed spikes with SMBHs $\lesssim10^7{\rm\,M}_\odot$, mass segregation may steepen the spike (to $\alpha\simeq2$) and enhance the merger rate, as long as the spike mass is dominated by lighter stars\,\citep{Alexander:2008tq,Preto:2009kd,Aharon:2016aa}.
In this case, PBHs may further dissipate their energy and angular momentum by scattering with stars, enhancing merger rates for EMRI and binary formation. Another possible enhancement of the EMRI rate in a relaxed spike is through SMBH binary mergers\,\citep{Mazzolari:2022cho,Naoz:2022rru,Naoz:2023hpz}.
We leave this to future work. 

\section{Direct collisions/plunges and merging timescales}
For mergers to be detectable by LIGO or LISA, they cannot be direct collisions (PBH binaries) or direct plunges (EMRIs). 
It is thus required that $r_{\rm min}$ in Eq.\,\ref{eq:closest} should be larger than the Schwarzschild radius of PBHs or SMBHs. The fraction of PBH--PBH direct collisions scales like $\propto(v/c)^{4/7}$, which is less than $1\%$ for $v\lesssim240{\rm\,km/s}$. Therefore, direct collisions are negligible. Interestingly, the fraction of PBH--SMBH direct plunges $\propto(M/m)^{2/7}(v/c)^{4/7}$ is larger than $93\%$ for $v\gtrsim240{\rm\,km/s}$ and $M\gtrsim10^8{\rm\,M}_\odot$, implying that sufficiently massive SMBHs do not produce EMRIs, but direct plunges. However, once the spin of the SMBH is accounted for, the majority of plunging orbits again become EMRIs\,\citep{Amaro-Seoane:2012jcd}.

We also require that the time for binaries to merge via GW emission is less than the Hubble time, which can be computed from $r_{\rm min}$, the periapsis of the newly formed bound orbits, in Eq.\,\ref{eq:closest}\,\citep{OLeary:2008myb}. The characteristic merger time is a function of the average initial separation and thus the PBH velocity dispersion in the spike. 
For PBH binaries, the merger time ranges from a few days to a few months. 
For EMRIs, the merger time is much longer, $10^3\textup{--}10^4{\rm\,yr}$. These timescales are all substantially less than the Hubble time. In addition, it can be easily checked that the PBH number density has to be much higher than that found in the spike for the disruption time to be comparable to their merging time. Three-body effects can thus be ignored for both PBH binaries and EMRIs.

Mergers from two-body relaxation can have a wider range of initial $r_{\rm min}$, and thus their merging times may be much larger than GW-driven mergers. If the merger time is larger than the Hubble time, such binaries would not be detected until GW emission circularizes their orbits. 
For EMRIs, it has also been shown that only stars or black holes originating from tightly bound orbits with semimajor axes $\lesssim10^{-2}{\rm\,pc}$ for $M\sim10^6{\rm\,M}_\odot$, can complete their inspiral without being scattered prematurely into the SMBH or onto a wider orbit\,\citep{Hopman:2006qr,Merritt:2011ve}.

%%%
\begin{figure}[t]
\centering
   \includegraphics[width=0.47\textwidth]{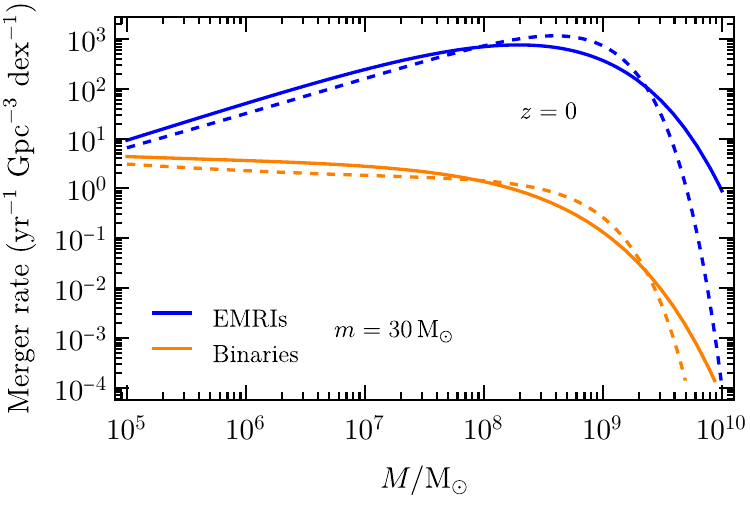}
   \caption{EMRI ({blue}) and PBH binary ({orange}) merger rates convolved with SMBH mass functions as a function of SMBH mass at redshift $z=0$. The solid lines denote the convolution using kinematic and photometric data based on the empirical relation between the halo velocity dispersion and the SMBH mass\,\citep{Shankar:2004ir}, while the dashed lines denote the convolution using the spheroid luminosity based on the assumption that all spheroids contain SMBHs at their centers\,\citep{Vika:2009aa}.}
   \label{fig:total_rate}
\end{figure}
%%%

\section{Expected events from LIGO and LISA}
To determine the expected LIGO and LISA events, we convolve the merger rates with the SMBH mass function ${\rm d}n/{\rm d}\log M(M)$.
The total merger rate per unit comoving volume is
\begin{equation}
\mathcal{V}_{\{\mathcal{R},\Gamma\}}
=\int\frac{{\rm d}n}{{\rm d}\log M}(M)
\left\{\begin{array}{lr}
\mathcal{R}(M)\\
\Gamma(M)
\end{array}\right\}
{\rm d}\log M.
\label{eq:total_rate}
\end{equation}

Fig.\,\ref{fig:total_rate} shows the convolved EMRI (blue) and PBH binary (orange) merger rates. We adopt two example SMBH mass functions from local Universe measurements for comparison: one using kinematic and photometric data\,\citep{Shankar:2004ir} and the other using the spheroid luminosity\,\citep{Vika:2009aa}. Their differences are moderate. The merger rates of EMRIs and PBH binaries are about the same at $M\sim10^5{\rm\,M}_\odot$, however, the EMRI rate increases and reaches a maximum of $\sim10^3{\rm\,yr^{-1}\,Gpc^{-3}\,dex^{-1}}$ at $M=10^{8}\textup{--}10^{9}{\rm\,M}_\odot$. The average estimated merger rate in the spike is $\mathcal{V}_\mathcal{R}\sim7{\rm\,yr^{-1}\,Gpc^{-3}}$\,(PBH binaries), while $\mathcal{V}_\Gamma\sim1.5\times10^3{\rm\,yr^{-1}\,Gpc^{-3}}$\,(EMRIs).

Generally, $\mathcal{V}_\mathcal{R}\propto f_{\rm PBH}^2$ while $\mathcal{V}_\Gamma\propto f_{\rm PBH}$ for GW-driven binary formation. On the other hand, using the number density $n={\rm d}N/{\rm d}V$ in Eq.\,\ref{eq:total_rate}, we can calculate the number of mergers per unit redshift $z$ expected over the observation time $T_{\rm obs}$ for LIGO or LISA as
\begin{equation}
\frac{{\rm d}N_{\{\mathcal{R},\Gamma\}}}{{\rm d}z}(z)=T_{\rm obs}\times\frac{\mathcal{V}_{\{\mathcal{R},\Gamma\}}(z)}{1+z}\frac{{\rm d}V}{{\rm d}z},
\label{eq:total_event}
\end{equation}
where ${\rm d}V/{\rm d}z$ is the comoving volume per unit redshift, and the $(1+z)^{-1}$ factor accounts for cosmological time dilation.
As LIGO and LISA are most sensitive to PBH binaries and EMRIs for $z\lesssim1$, respectively, and the change in the rates is negligible from $z=0$ to $1$ assuming the same SMBH mass functions, we can ignore redshift evolution. The comoving volume is approximately $\approx165{\rm\,Gpc^3}$ for $z\lesssim1$, which corresponds to approximately $\sim1.5\times10^5$ EMRI events associated with $\sim7\times10^2$ PBH binaries per year with $f_{\rm PBH}=1$, while we expect $\sim1.5\times10^3$ EMRIs associated with $<1$ PBH binary per year with $f_{\rm PBH}=0.01$.

%%%
\begin{table}[t]
\center
\begin{tabular}{|p{2.6cm}||p{2.3cm}|p{2.3cm}|}
 \hline
  \multicolumn{3}{|c|}{EMRI Merger Rates ${\rm\,(yr^{-1}\,Gpc^{-3}})$} \\
 \hline
 PBH Mass $({\rm M}_\odot)$ & Shankar04 & Vika09 \\
 \hline
 1 & $1.55314\times10^{4}$ & $1.84203\times10^{4}$ \\
 3 & $7.08615\times10^{3}$ & $8.40420\times10^{3}$ \\
 10 & $2.99865\times10^{3}$ & $3.55641\times10^{3}$ \\
 30 & $1.36129\times10^{3}$ & $1.61796\times10^{3}$ \\
 100 & $5.68596\times10^{2}$ & $6.79906\times10^{2}$ \\
 \hline
\end{tabular}
\caption{The EMRI merger rates for $1,3,10, 30$, and $100{\rm\,M}_\odot$ PBHs, calculated using SMBH functions from~\cite{Shankar:2004ir} and \cite{Vika:2009aa}, assuming an inner cusp index $\gamma=1.7$ to match the DM spike described by the Bahcall--Wolf power law ($\alpha=7/4$).}
\label{tab:EMRI_rates}
\end{table}

%%%
\begin{figure}[t]
\centering
   \includegraphics[width=0.47\textwidth]{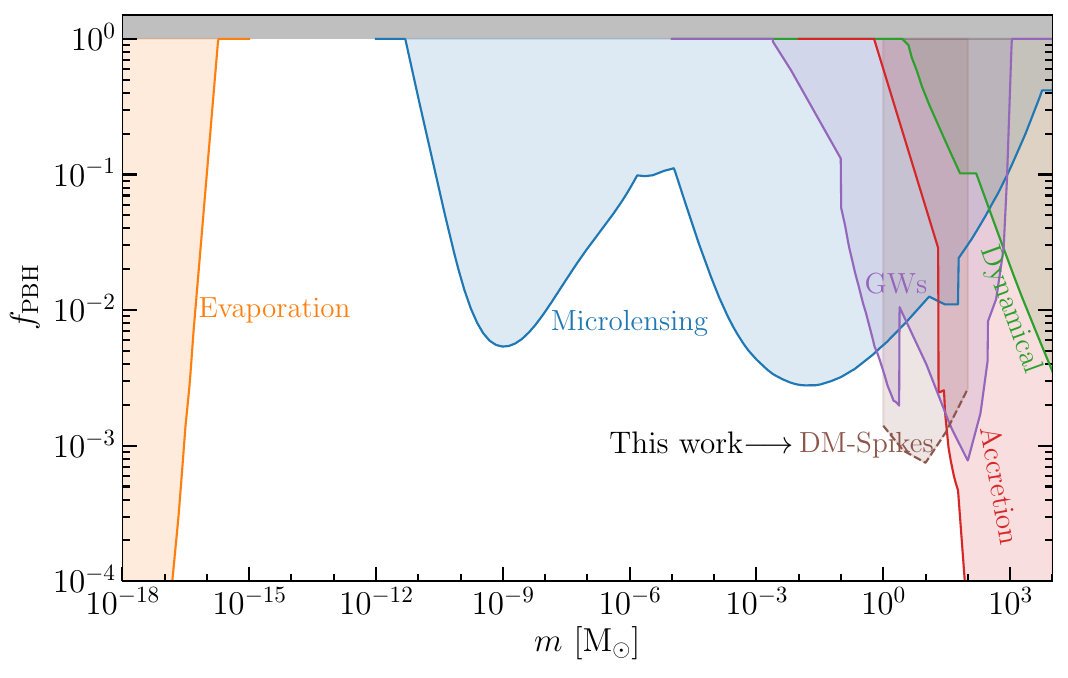}
   \caption{PBH fraction $f_{\rm PBH}$ in DM as a function of PBH mass $m$, assuming a monochromatic mass spectrum. Forecast constraints from EMRIs in DM spikes are given for the mass range $\sim1\textup{--}100{\rm\,M}_\odot$, assuming that 20 events are detected during LISA's 4\,yr observation run with a detection threshold $\varrho_{\rm th}=20$. The current constraints for a broader mass range are also shown; from~\cite{PBHbounds} and the references therein.}
   \label{fig:bounds}
\end{figure}
%%%

So far our results focus on $30{\rm\,M}_\odot$ PBHs, and the binary merger rate is only marginally sensitive to the assumed mass. However, the EMRI rate is sensitive to the PBH mass. As indicated by the scaling relation in Eq.\,\ref{eq:emri_rate_est}, higher EMRI rates are expected for lower PBH masses, and vice versa. In Table\,\ref{tab:EMRI_rates}, we consider the PBH masses ranging from $1$ to $100{\rm\,M}_\odot$, motivated by their formation during the QCD phase transition\,\citep{Musco:2023dak} and falling within the LVK detection range for black hole binaries\,\citep{KAGRA:2021duu}. The merger rate is calculated as the area under Fig.\,\ref{fig:total_rate} for EMRIs with $M/m>10^4$. The expected EMRI rate is as high as $\sim1.7\times10^4{\rm\,yr^{-1}\,Gpc^{-3}}$ for PBHs of $1{\rm\,M}_\odot$; while as low as $\sim6\times10^2{\rm\,yr^{-1}\,Gpc^{-3}}$ for $100{\rm\,M}_\odot$ PBHs. Similarly, we can estimate the expected event number as before following Eq.\,\ref{eq:total_event}, while attached with a step function $\theta(\varrho-\varrho_{\rm th})$ to extract EMRIs with the required detection threshold $\varrho_{\rm th}$, and the signal-to-noise ratio $\varrho$ is given by\,\citep{Moore:2014lga}
\begin{equation}
\varrho^2\equiv4\int_0^{f_{\rm max}}\frac{|\tilde{h}(f)|^2}{S_n(f)}{\rm\,d}f
\label{eq:snr}
\end{equation}
for a GW signal whose Fourier transform is $\tilde{h}(f)$ in the LISA detector of noise spectral density $S_n(f)$\,\cite[][see also Appendix E]{Babak:2021mhe}, where $f_{\rm max}$ is given by the frequency at the innermost stable circular orbit:
\begin{equation}
f_{\rm ISCO}\simeq4.7\times10^{-3}{\rm\,Hz}\left(\frac{1}{1+z}\right)\left(\frac{10^6{\rm\,M}_\odot}{M+m}\right).
\end{equation}
The number of resolved EMRIs, divided by the expected number, provides an upper bound on $f_{\rm PBH}$ in the spike. 

Fig.\,\ref{fig:bounds} shows the forecast constraints on the PBH DM fraction in the mass range $\sim1\textup{--}100{\rm\,M}_\odot$, where we assume $20$ resolved EMRI events are observed for LISA's 4\,yr observation with a threshold $\varrho_{\rm th}=20$\,\citep{Gair:2017ynp}, instead of fixing $f_{\rm PBH}=1$. 
Specifically, we obtain $\sim(1.4,\,2.2,\,2.7,\,1.6)\times10^4$ and $7.6\times10^3$ expected EMRIs for $(1,\,3,\,10,\,30)$ and $100{\rm\,M}_\odot$ PBHs, respectively, out to $z=1$. Interestingly, although the theoretical rate is higher for smaller PBHs as shown in Table\,\ref{tab:EMRI_rates}, the low chirp mass $\mathcal{M}\sim m^{3/5}M^{2/5}$ reduces their detectability $\varrho^2\propto \mathcal{M}^{5/3}\sim mM^{2/3}$, such that the expected events increase with PBH mass and peak around $\sim10{\rm\,M}_\odot$ then decrease. Overall, this places much stronger constraints than current microlensing surveys: $f_{\rm PBH}\sim10^{-4}\textup{--}10^{-3}$. These constraints are conservative as we assume all the events are from PBHs. In reality, EMRIs could also form when a stellar black hole from a star cluster spirals into an SMBH. If EMRIs could confidently be ascribed to a non-PBH source, then constraints on the PBH abundance in the spike would be stronger. 

While unresolved EMRIs will contribute to confusion noise in the stochastic GW background, the merger frequencies of these EMRIs are not observable at plausible signal-to-noise in a midband experiment\,\citep{Barish:2020vmy} or the nano-Hertz band by pulsar timing array experiments\,\citep{NANOGrav:2023gor,Reardon:2023gzh,EPTA:2023fyk,Xu:2023wog} due to their low GW emissivity at these frequencies. Therefore, LISA remains the most promising facility for constraining PBHs.

\section{Discussion and conclusions} 
We have estimated the merger rates in the spike at redshift $z=0$ to be $\mathcal{V}_\mathcal{R}\simeq7{\rm\,yr^{-1}\,Gpc^{-3}}$ for PBH binaries, and $\mathcal{V}_\Gamma\simeq1.5\times10^3{\rm\,yr^{-1}\,Gpc^{-3}}$ for EMRIs, assuming a PBH DM fraction of $f_{\rm PBH}=1$. Although we have assumed a single population of PBHs of mass $\sim30{\rm\,M}_\odot$, our results should be largely insensitive to a small spread in the mass function around this nominal value.
Interestingly, the recent LVK estimation of two merging black holes, $m_1\in[5, 50]{\rm\,M}_\odot$ and $m_2\in[20, 50]{\rm\,M}_\odot$ implies an event rate of $2.5\textup{--}6.3{\rm\,yr^{-1}\,Gpc^{-3}}$ within the $90\%$ credible range\,\citep{KAGRA:2021duu}. As this is less than our estimates resulting from PBH DM spikes, it suggests $f_{\rm PBH} \lesssim 1$ in the spike. 

The EMRI mergers arising in the spike could be observed with the future LISA mission\,\citep{Colpi:2024xhw}. As LISA is most sensitive to EMRI systems with mass $M\simeq10^4\textup{--}10^7{\rm\,M}_\odot$, the expected EMRI rates for $M\gtrsim10^7{\rm\,M}_\odot$ may be below the detection threshold. However, we showed that LISA EMRI rates can place strong constraints on the fraction of DM made up of PBHs. Importantly, the EMRI rate scales as $f_{\rm PBH}$, rather than $f_{\rm PBH}^2$ for the binary merger rate, and so retains sensitivity even to low PBH DM fractions. Black hole binary mergers from LVK could also complement this mass range given the correlation $\Gamma/\mathcal{R}\propto M^{5/7}m^{-5/7}$. EMRI and PBH merger rates will be spatially correlated, and this correlation may be detectable with more sensitive next-generation ground-based GW experiments, such as the Einstein Telescope\,\citep{Coccia:2023wag} and Cosmic Explorer\,\citep{Reitze:2019iox}, together with space-based GW detectors DECIGO\,\citep{Kawamura:2020pcg}, LISA, Taiji, and Tianqin\,\citep{Gong:2021gvw}.

\section*{Acknowledgements}
We thank the anonymous referee for improving the draft.
WXF acknowledges support from the Anne Kernan Award, Dissertation Year Program Award, and Tsinghua's Shuimu Scholar Fellow during the completion of this work. This work was supported in part by the China Postdoctoral Science Foundation under Grant No.\,2024M761594\,(WXF), NASA ATP 80NSSC22K1897\,(SB), the John Templeton Foundation under Grant ID\#61884, and the U.S. Department of Energy under Grant No.\,DE-SC0008541\,(HBY). The opinions expressed in this publication are those of the authors and do not necessarily reflect the views of the funding agencies.

\appendix

\section{SMBH mass as a function of halo mass}
The merger rate generally depends on redshift $z$ through the scale density and radius of NFW, given by~\cite{Navarro:1995iw}:
\begin{equation}
\rho_s(z)=\frac{200c_{200}^3\rho_c(z)}{3f(c_{200})}
\quad{\rm and}\quad r_s(z)=\left(\frac{3M_{200}}{800\pi c_{200}^3\rho_c(z)}\right)^{1/3}
\end{equation}
with
\begin{equation}
\rho_c(z)=\frac{3H_0^2}{8\pi G}\left[\Omega_{M,0}(1+z)^3+\Omega_\Lambda\right],
\end{equation}
$H_0=1.02\times10^{-4}h{\rm\,Myr}^{-1}$, $h=0.67556$, $\Omega_{M,0}=0.312$, and $\Omega_\Lambda=0.688$\,\citep{Planck:2018vyg}.

\begin{figure}[b]
\centering
   \includegraphics[width=0.47\textwidth]{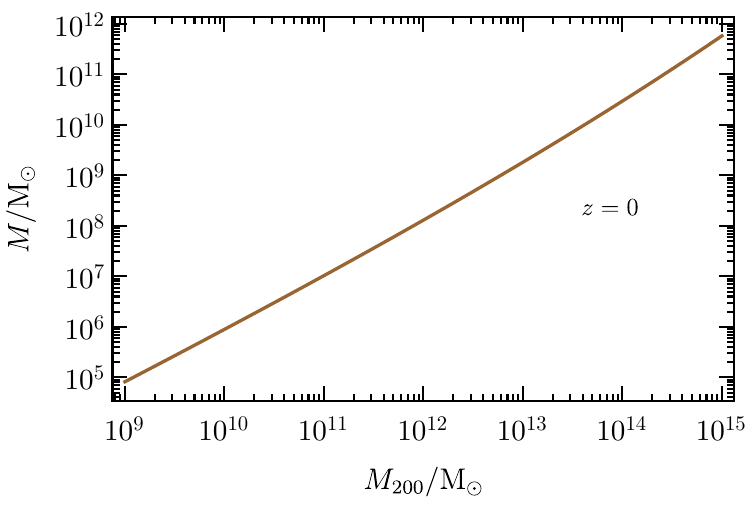}
   \caption{SMBH mass as a function of halo mass at redshift $z=0$, following the $M\textup{--}\sigma_*$ and concentration--mass relations.}
   \label{fig:BH_mass}
\end{figure}

We adopt the $M\textup{--}\sigma_*$ relation\,\citep{Ferrarese:2000se,Gebhardt:2000fk,Tremaine:2002js} 
\begin{equation}
\log\left(M/{\rm M}_\odot\right)=a_*+b_*\log\left(\sigma_*/200{\rm\,km\,s^{-1}}\right)
\end{equation}
with the parameters $a_*=8.12\pm0.08$ and $b_*=4.24\pm0.41$\,\citep{Gultekin:2009aa,Kormendy:2013dxa}, where we identify
\begin{equation}
\sigma_*=\frac{1}{\sqrt{3}}v_{\rm dm}=\sqrt{\frac{GM_{\rm max}}{R_{\rm max}}}
=\sqrt{\frac{4\pi G\rho_s r_s^2f(c_{\rm max})}{c_{\rm max}}}.
\end{equation}
To determine the SMBH mass and merger rate, however, we need the concentration--mass ($c_{200}\textup{--}M_{200}$) relation as a function of redshift\,\citep{Prada:2011aa,Ludlow:2013vxa,Correa:2015dva}. 
We use the following relation\,\citep{Prada:2011aa,Loudas:2022ipe}: 
\begin{equation}
\log c_{200}=4.23-0.25\log(M_{200}/{\rm M}_\odot)-0.16\log\left(\frac{1+z}{1.47}\right).
\end{equation} 
In Fig.\,\ref{fig:BH_mass}, we plot the SMBH mass as a function of halo mass at redshift $z=0$.
%%%
%%%

\section{Matching the spike to the NFW cusp}
We need to match the spike and NFW profiles.
In particular, the total DM spike mass $M_{\rm sp}$ and hence $\Delta$ is sensitive to the inner cusp index $\gamma$, which is not necessarily $1$, near the spike. We then consider the \emph{modified} NFW profile $\rho_{\rm NFW}(r)=\rho_s(r/r_s)^{-\gamma}(1+r/r_s)^{-2}\simeq\rho_s(r/r_s)^{-\gamma}$ when $r\ll r_s$, where the cusp index $\gamma$ ranges from $0$ to $2$\,\citep{Ostriker:1999ee,deBlok:2009sp,Shi:2021tyg}.
In other words,  one can determine $\Delta=\Delta(\alpha, \beta, M, M_{200}, c_{200})$ through matching the modified NFW profile as $r_{\rm sp}\ll r_s$, such that 
\[
\rho_{\rm sp}=\rho_{\rm NFW}(r_{\rm sp})=\rho_s(r_{\rm sp}/r_s)^{-\gamma}
\]
\begin{equation}
=\rho_s\left(\frac{2M}{3M_{200}}f(c_{200})\frac{c_{\rm max}}{f(c_{\rm max})}\right)^{-\gamma},
\label{eq:rhosp1}
\end{equation}
where we have used the Bondi radius as the spike radius\,\footnote{Technically, the Bondi radius $r_{\rm B}=2GM/c_s^2$ is defined through the ``sound speed'' $c_s$ of the accreted gas far away from the Bondi radius\,\citep{Armitage:2020owb}. For PBHs as DM, the ``effective'' $c_s^2=(5/9)v_{\rm dm}^2$, thus the spike radius $r_{\rm sp}=2GM/v_{\rm dm}^2$ is smaller than $r_{\rm B}$.}
\[
r_{\rm sp}=\frac{2GM}{v_{\rm dm}^2}=\frac{2M}{3M_{\rm max}}R_{\rm max}
\]
\begin{equation}
=\frac{2M}{3M_{200}}f(c_{200})\frac{c_{\rm max}}{f(c_{\rm max})}r_s.
\end{equation}
Together with the ansatz, $M_{\rm sp}=\Delta M\simeq4\pi\rho_{\rm sp}r_{\rm sp}^3/(3-\alpha)$, we have (see also~\cite{Feng:2023jvn})
\begin{equation}
\rho_{\rm sp}=\frac{27}{8}(3-\alpha)\Delta\rho_s\left(\frac{M_{200}}{M}\right)^2\frac{1}{f^2(c_{200})}\left[\frac{f(c_{\rm max})}{c_{\rm max}}\right]^3.
\label{eq:rhosp2}
\end{equation}
Eqs.\,\ref{eq:rhosp1} and \ref{eq:rhosp2} give $\Delta=\Delta(\alpha, \gamma, M, M_{200}, c_{200})$
\[
=\frac{8/27}{3-\alpha}\left(\frac{M}{M_{200}}\right)^2f^2(c_{200})\left[\frac{c_{\rm max}}{f(c_{\rm max})}\right]^3
\]
\begin{equation}
\times\left(\frac{2M}{3M_{200}}f(c_{200})\frac{c_{\rm max}}{f(c_{\rm max})}\right)^{-\gamma},
\end{equation}
while with the $M\textup{--}\sigma_*$ and concentration--mass relations, $\Delta=\Delta(\alpha, \gamma, M)$ depends only on the black hole mass $M$, the spike index $\alpha$ and the cusp index $\gamma$. 

\begin{figure}[!t]
   \includegraphics[width=0.48\textwidth]{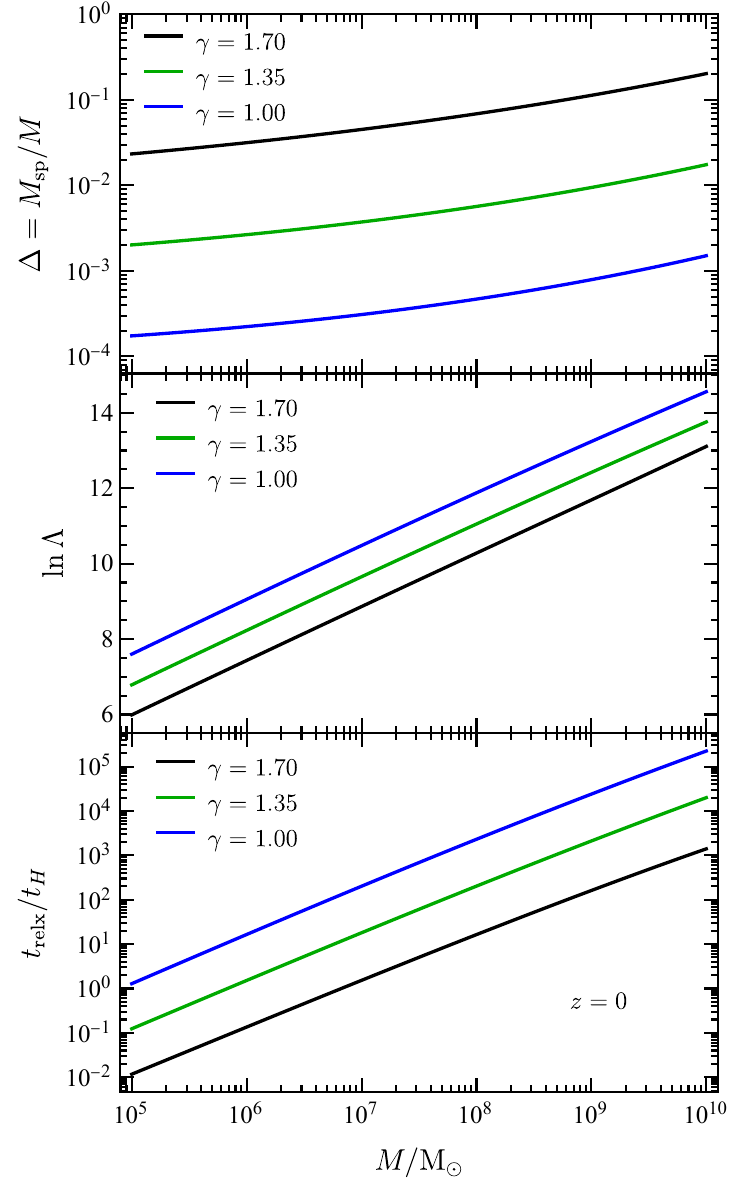}
   \caption{The ratio of spike to SMBH masses (top panel), Coulomb logarithm (middle panel), ratio of relaxation time to the Hubble time (bottom panel) at redshift $z=0$, following the $M\textup{--}\sigma_*$ and concentration--mass relations. We assume Bahcall--Wolf power law $\alpha=7/4$ for inner cusp indices $\gamma=1.7, 1.35$, and $1$. The PBH mass is $30{\rm\,M}_\odot$.}
   \label{fig:delta_coulomb_relax}
\end{figure}

Fig.\,\ref{fig:delta_coulomb_relax} shows the DM spike fraction $\Delta=M_{\rm sp}/M$ ({top panel}), the corresponding Coulomb logarithm $\ln\Lambda$  ({middle panel}), and relaxation time to Hubble time ratio $t_{\rm relx}/t_H$  ({bottom panel}). It turns out $\gamma=1$ gives $\Delta\simeq0.0008$; $\gamma=1.35$ gives $\Delta\simeq0.009$; $\gamma=1.7$ gives $\Delta\simeq0.1$ at $M=10^{9}{\rm\,M}_\odot$ with Bahcall--Wolf spike $\alpha=7/4$. 

To compare another definition of $r_{\rm sp}$ in the literature, we rewrite 
\[
M_{\rm sp}\simeq\frac{4\pi\rho_{\rm sp}r_{\rm sp}^3}{3-\alpha}
\simeq\frac{4\pi\rho_s\left(r_{\rm sp}/r_s\right)^{-\gamma}r_{\rm sp}^3}{3-\alpha}
=\frac{4\pi\rho_s r_s^\gamma r_{\rm sp}^{3-\gamma}}{3-\alpha}
\]
hence
\[
r_{\rm sp}=\left[\frac{3-\alpha}{4\pi}\frac{M_{\rm sp}}{\rho_sr_s^\gamma}\right]^{\frac{1}{3-\gamma}}
\]
\begin{equation}
=r_s\left[\frac{3-\alpha}{4\pi}\frac{M_{\rm sp}}{\rho_sr_s^3}\right]^{\frac{1}{3-\gamma}}
=r_s\left[\frac{3-\alpha}{4\pi}\frac{\Delta M}{\rho_sr_s^3}\right]^{\frac{1}{3-\gamma}}.
\end{equation}
In the literature\,\citep{Merritt:2003qc,Lenoci:2023gjz}, a commonly used definition for $r_{\rm sp}$ is through the enclosed mass of DM within the ``radius of influence'' $r_h$ equal to 2 times the central SMBH mass, i.e.,  $M_{\rm dm}(r<r_h)=2M$ and the spike extends up to $r_{\rm sp}\simeq0.2r_h$. This implies
\[
2M=\frac{4\pi\rho_{\rm sp}(r_{\rm sp}/0.2)^{3-\alpha}}{3-\alpha}r_{\rm sp}^\alpha
\]
thus
\[
r_{\rm sp}=\left[\frac{(0.2)^{3-\alpha}(3-\alpha)M}{2\pi\rho_{\rm sp}}\right]^{1/3}
=\left[\frac{(0.2)^{3-\alpha}(3-\alpha)M}{2\pi\rho_s(r_{\rm sp}/r_s)^{-\gamma}}\right]^{1/3}.
\]
Solving for $r_{\rm sp}$, we obtain
\begin{equation}
r_{\rm sp}=r_s\left[\frac{(0.2)^{3-\alpha}(3-\alpha)M}{2\pi\rho_sr_s^3}\right]^{\frac{1}{3-\gamma}}.
\end{equation}
In comparison, this definition \emph{automatically} fixes the spike fraction to be 
$\Delta=\Delta(\alpha)=2\times(0.2)^{3-\alpha}=0.267$
for spike index $\alpha=1.75$, independent of the halo properties and the black hole mass.
%%%
%%%

\section{Merger rates driven by GW emission}
The velocity-weighted cross sections for PBH mergers and EMRIs are
\begin{equation}
\langle\sigma_{\rm merg}^{\rm PBH} v\rangle
=(4\pi)^2\left(\frac{85\pi}{3}\right)^{2/7}\frac{G^2m^2}{c^3}D(\sigma_{\rm sp})
\end{equation}
and
\begin{equation}
\langle\sigma_{\rm merg}^{\rm EMRI} v\rangle
=(4\pi)^2\left(\frac{85\pi}{96}\right)^{2/7}\left(\frac{M}{m}\right)^{12/7}\frac{G^2m^2}{c^3}D(\sigma_{\rm sp}),
\end{equation}
respectively,
where
\begin{equation}
D(\sigma_{\rm sp})=\int_0^{v_{\rm cut}}c^3\mathcal{P}(v,\sigma_{\rm sp})\left(\frac{v}{c}\right)^{3/7}{\rm d}\left(\frac{v}{c}\right).
\end{equation}

Then the resulting merger rates for PBH binaries and EMRIs are
\[
\left\{\begin{array}{lr}
\mathcal{R}\\
\Gamma
\end{array}\right\}
=
\left\{\begin{array}{lr}
\mathcal{C}_{\rm PBH}\\
\mathcal{C}_{\rm EMRI}
\end{array}\right\}
\frac{G^2\rho_sM}{c^3}\left(\frac{M_{200}}{M}\right)^2
\]
\begin{equation}
\times\frac{D(\sigma_{\rm sp})}{f^2(c_{200})}\left[\frac{f(c_{\rm max})}{c_{\rm max}}\right]^3
\end{equation}
with 
\[
\mathcal{C}_{\rm PBH}=27\pi^2\left(\frac{85\pi}{3}\right)^{2/7}(3-\alpha)^2\Delta^2
\]
\begin{equation}
\times
\begin{cases}
\left(\frac{1}{3-2\alpha}\right)
\left[1-\left(v_{\rm dm}^2/c^2\right)^{3-2\alpha}\right]~\left(\alpha\neq3/2\right) 
\\
\ln\left(c^2/{v_{\rm dm}^2}\right)~\left(\alpha=3/2\right)
\end{cases}
\end{equation}
and
\begin{equation}
\mathcal{C}_{\rm EMRI}=162\pi^2\left(\frac{85\pi}{96}\right)^{2/7}\left(\frac{M}{m}\right)^{5/7}\Delta,
\end{equation}
respectively. We set $v_{\rm dm}^2=3GM_{\rm max}/R_{\rm max}=12\pi G\rho_sr_s^2f(c_{\rm max})/c_{\rm max}$ to associate with the halo properties.

\section{The SMBH mass functions and merger rates}
%%%
\begin{figure}[t]
\centering
   \includegraphics[width=0.47\textwidth]{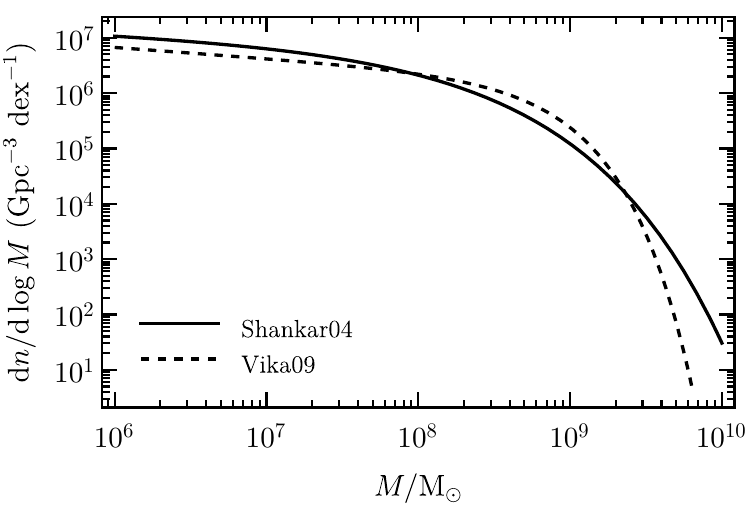}
   \caption{Estimated SMBH mass functions using kinematic and photometric data based on the empirical relation between the halo velocity dispersion and the SMBH mass in~\cite{Shankar:2004ir}, and using the spheroid luminosity based on the assumption that all spheroids contain SMBHs at their centers in~\cite{Vika:2009aa}.}
   \label{fig:BH_mass_func}
\end{figure}
%%%
We consider two SMBH mass functions.
First, the estimated SMBH mass function according to kinematic and photometric data based on the empirical relation between the halo velocity dispersion and the SMBH mass in~\cite{Shankar:2004ir},
\begin{equation}
\frac{{\rm d}n}{{\rm d}\log M}(M)=\phi_*\left(\frac{M}{M_*}\right)^{A+1}\exp\left[-\left(\frac{M}{M_*}\right)^B\right]
\end{equation}
with $\phi_*=7.7\times10^{6}{\rm\,Gpc^{-3}}$, $M_*=6.4\times10^7{\rm\,M}_\odot$, $A=-1.11$ and $B=0.49$.
The other one is based on the assumption that all spheroids contain SMBHs at their centers, using 1743 galaxies from the Millennium Galaxy Catalog in~\cite{Vika:2009aa},
\begin{equation}
\frac{{\rm d}n}{{\rm d}\log M}(M)=\phi_*\left(\frac{M}{M_*}\right)^{A+1}\exp\left[1-\left(\frac{M}{M_*}\right)\right]
\end{equation}
with $\phi_*=7.07946\times10^5{\rm\,Gpc^{-3}}$, $M_*=10^{8.71}{\rm\,M}_\odot$ and $A=-1.2$. See Fig.\,\ref{fig:BH_mass_func} for comparison of the two mass functions and Table\,\ref{tab:merger_rates} for the corresponding PBH binary and EMRI merger rates of $30{\rm\,M}_\odot$ PBHs with inner cusp indices $\gamma=1.7, 1.35,$ and $1$.

%%%
\begin{table}[t]
\center
\begin{tabular}{|p{2.8cm}||p{2.3cm}|p{2.3cm}|}
 \hline
 \multicolumn{3}{|c|}{PBH Binary Merger Rates ${\rm\,(yr^{-1}\,Gpc^{-3}})$} \\
 \hline
 Inner Cusp Index $\gamma$ & Shankar04 & Vika09 \\
 \hline
 1.70 & $8.00776\times10^{0}$ & $5.88495\times10^{0}$  \\
 1.35 & $5.73483\times10^{-2}$ & $4.21906\times10^{-2}$  \\
 1.00 & $3.99449\times10^{-4}$ & $2.93342\times10^{-4}$  \\
 \hline
 \hline
  \multicolumn{3}{|c|}{EMRI Merger Rates ${\rm\,(yr^{-1}\,Gpc^{-3}})$} \\
 \hline
 Inner Cusp Index $\gamma$ & Shankar04 & Vika09 \\
 \hline
 1.70 & $1.36129\times10^{3}$ & $1.61796\times10^{3}$ \\
 1.35 & $1.18750\times10^{2}$ & $1.41930\times10^{2}$ \\
 1.00 & $9.88148\times10^{0}$ & $1.18237\times10^{1}$ \\
 \hline
\end{tabular}
\caption{The merger rates of $30{\rm\,M}_\odot$ PBH binaries and EMRIs with various inner cusp indices $\gamma=1.7, 1.35,$ and $1$ to match the DM spike of Bahcall--Wolf power law $\alpha=7/4$, given the SMBH functions in~\cite{Shankar:2004ir} and \cite{Vika:2009aa}.}
\label{tab:merger_rates}
\end{table}

In the main text, we assume $\Delta\simeq0.1$ (of all PBH DM) at $M=10^9{\rm\,M}_\odot$ but vary with the SMBH mass, which fixes the inner cusp index of all halos to be $\gamma=1.7$. Assuming $\Delta=0.1$ uniformly over the whole SMBH spectrum will result in more enhancement of $\mathcal{V}_\mathcal{R}\sim35\textup{--}53{\rm\,yr^{-1}\,Gpc^{-3}}$ (PBH binaries), while $\mathcal{V}_\Gamma\sim2\times10^3{\rm\,yr^{-1}\,Gpc^{-3}}$ (EMRIs), from GW-driven processes.
Moreover, if we adopt the spike radius $r_{\rm sp}$ by setting the enclosed DM mass $M_{\rm dm}(r<r_{\rm sp}/0.2)=2M$\,\citep{Merritt:2003qc,Lenoci:2023gjz}, the DM fraction $\Delta=0.267$ is fixed and much more events are expected. In this regard, our estimated rates shown in the main text are conservative.

\section{LISA sensitivity and strain amplitude}
We adopt the official required sensitivity of LISA to calculate Eq.\,\ref{eq:snr}:
\begin{align}
S_h(f)&=\frac{1}{2}\frac{20}{3}\left(\frac{S_1(f)}{(2\pi f)^4}+S_2(f)\right)S_3(f)\,,\notag\\
S_1(f)&=5.76\times10^{-48}\left[1+(f_1/f)^2\right]{\rm\,s^{-4}\,Hz^{-1}}\,,\notag\\
S_2(f)&=3.6\times10^{-41}{\rm\,Hz^{-1}},\, S_3(f)=1+(f/f_2)^2\,,\notag\\
f_1&= 0.4{\rm\,mHz},{\rm\,and}\,\, f_2= 25{\rm\,mHz}\;.
\end{align} 
The strain amplitude, after the polarization and sky averaging, is given by\,\citep{Babak:2021mhe}
\begin{equation}
|\tilde{h}(f)|=\frac{1}{2\sqrt{10}}\frac{(G\mathcal{M}_z)^{5/6}}{\pi^{2/3}c^{3/2}d_L(z)}f^{-7/6}
\end{equation}
and the noise $S_n(f)$ in Eq.\,\ref{eq:snr} is related to $S_h(f)$ by $S_h=5S_n$, where $\mathcal{M}_z=(1+z)\mathcal{M}$ is the (redshifted) chirp mass in the detector frame with $\mathcal{M}^{5/3}\equiv mM/(m+M)^{1/3}$ in the source frame, and $d_L(z)=(1+z)R(z)$ is the luminosity distance with $R(z)$ the comoving distance.

\bibliography{pbh}{}
\bibliographystyle{aasjournal}

\end{document}